\documentclass[prl,superscriptaddress,amsmath,amssymb,citeautoscript,showpacs,reprint]{revtex4-1}
\usepackage[pass]{geometry}
\usepackage[utf8]{inputenc}
\usepackage[T1]{fontenc}
\usepackage[colorlinks=true,urlcolor=blue,linkcolor=blue,citecolor=blue]{hyperref}

\usepackage{graphicx}

\usepackage{pdfpages}

\makeatletter
\newcommand*{\balancecolsandclearpage}{%
  \close@column@grid
  \cleardoublepage
  \twocolumngrid
}
\makeatother
\newcommand{\V}[1]{\boldsymbol #1}
\newcommand{\dee}{\text{d}}

\renewcommand{\Im}{\text{Im}}

\newcommand\textlcsc[1]{\textsc{\MakeLowercase{#1}}}

\begin{document}

\title{Quantized evolution of the plasmonic response in a stretched nanorod}
\date{\today}
\author{Tuomas\ P.\ Rossi}
\email{tuomas.rossi@alumni.aalto.fi}

\affiliation{COMP Centre of Excellence, Department of Applied Physics, 
Aalto University School of Science, P.O.\ Box 11100, FI-00076 Aalto, 
Finland}

\author{Asier\ Zugarramurdi}
\affiliation{COMP Centre of Excellence, Department of Applied Physics, 
Aalto University School of Science, P.O.\ Box 11100, FI-00076 Aalto, 
Finland}

\author{Martti\ J.\ Puska}
\affiliation{COMP Centre of Excellence, Department of Applied Physics, 
Aalto University School of Science, P.O.\ Box 11100, FI-00076 Aalto, 
Finland}

\author{Risto\ M.\ Nieminen}
\affiliation{COMP Centre of Excellence, Department of Applied Physics, 
Aalto University School of Science, P.O.\ Box 11100, FI-00076 Aalto, 
Finland}
\affiliation{Dean's Office, Aalto University School of Science, 
P.O.\ Box 11000, FI-00076 Aalto, Finland}

\begin{abstract}
Quantum aspects, such as electron tunneling between closely separated
metallic nanoparticles, are crucial for understanding the plasmonic
response of nanoscale systems.  We explore quantum effects on the
response of the conductively coupled metallic nanoparticle dimer.
This is realized by stretching a nanorod, which leads to the formation
of a narrowing atomic contact between the two nanorod ends.  Based on
first-principles time-dependent density-functional-theory
calculations, we find a discontinuous evolution of the plasmonic
response as the nanorod is stretched. This is especially pronounced
for the intensity of the main charge-transfer plasmon mode. We show
the correlation between the observed discontinuities and the discrete
nature of the conduction channels supported by the formed atomic-sized
junction.
\end{abstract}


\maketitle

Metallic nanoparticles are of great interest because they have
potential in many applications, such as biosensor development
\cite{Anker2008}, spectroscopy \cite{Aizpurua2005}, and
nanobiomedicine \cite{Huang2009}.  This is thanks to their versatile
optical properties that are mainly determined by the localized surface
plasmon resonances supported by the nanoparticle. These excitations
are characterized by the collective oscillations of the delocalized
electrons in response to light. Understanding the physics determining
their dependence on nanoparticle parameters is thus a major subject of
research seeking for different routes to tailor materials with desired
optical properties.  Along with the size, shape, environment, and
atomic composition of the individual nanoparticles, the plasmonic
response can be modified by forming extended systems, such as
nanoparticle arrays and agglomerates \cite{Herrmann2014}. Among these,
nanoparticle dimers are not only the simplest configurations for
studying the plasmonic interparticle interaction, they also serve as
basic components in the design of nanoantennas 
\cite{Novotny2011,Georgiou2015}.

The plasmonic response of large nanoparticle dimers can be mostly
understood by classical electrodynamics and the plasmon hybridization
models \cite{Prodan2003,Nordlander2004}. However, the classical
description fails at subnanometer interparticle separations, where
quantum tunneling of electrons strongly affects the plasmonic
response. Recently, this \emph{quantum regime} has become
experimentally accessible \cite{Savage2012,Scholl2013} confirming
several theoretical predictions 
\cite{Zuloaga2009,Esteban2012,Marinica2012}.
Follow-up studies have further addressed quantum effects from dimers
of mesoscopic sizes \cite{Yan2015} down to the atomistic details
\cite{Zhang2014}, also considering the formation of conductive
contacts \cite{Varas2015,Barbry2015}, as well as molecular and atomic
junctions between the nanoparticles
\cite{Song2011,Song2012,Tan2014,Kulkarni2015}.

In this Letter we report, based on first-principles calculations, that
the plasmonic response of nanoparticle dimers shows important quantum
features within the metal-contact regime.  We consider a metallic
nanorod subject to a stretching process leading to the formation of a
narrowing atomic contact between the two nanorod ends until eventual
breakage.  Remarkably, the calculations reveal a discontinuous
evolution of the plasmon energies and intensities during stretching.
We find a strong correlation between the observed discontinuities and
the quantized nature of the conductance of the narrow contact. This
effect is pronounced for the main charge-transfer plasmon mode.

\begin{figure*}
 \centering
 \includegraphics[scale=1]{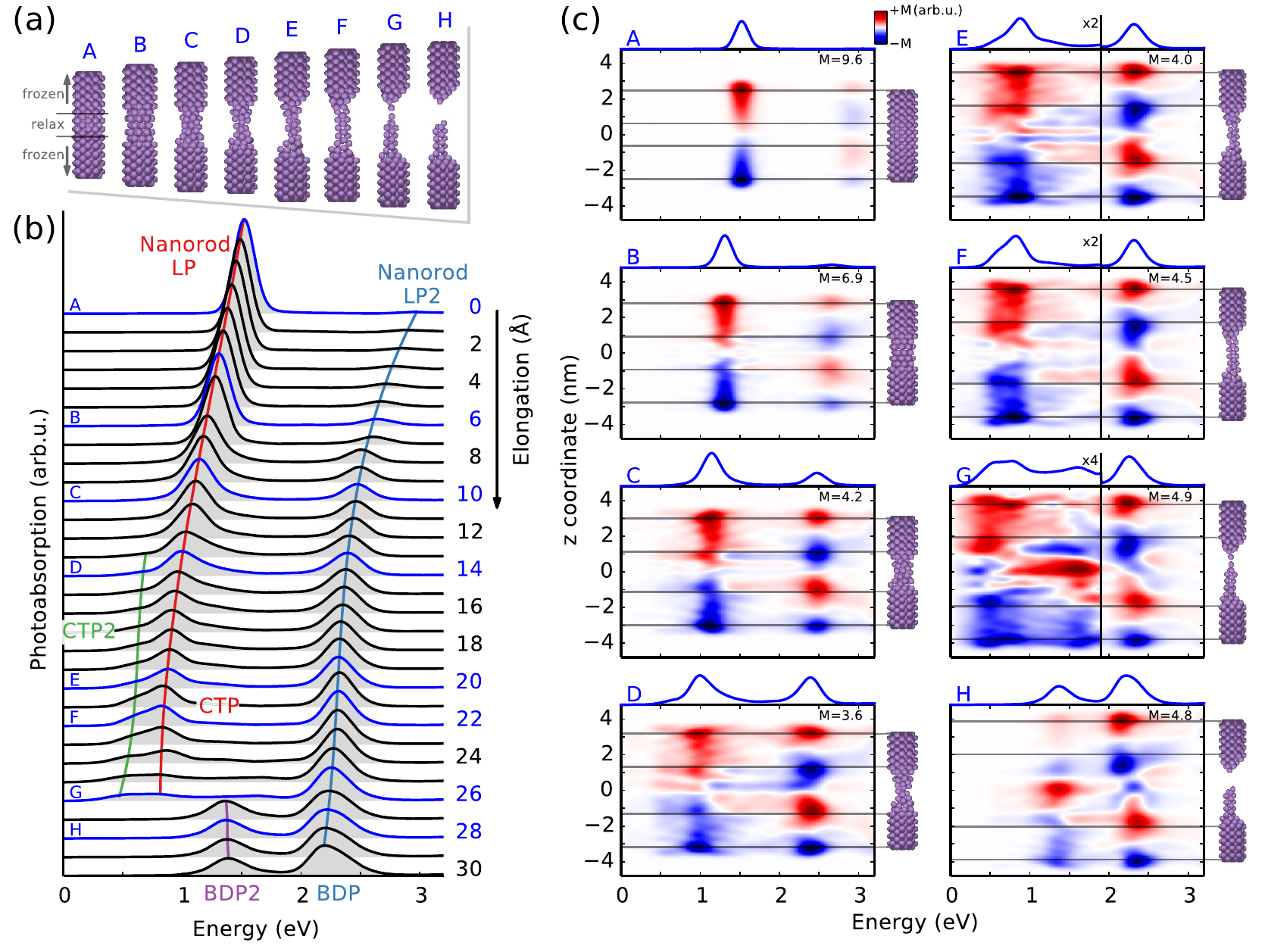}
 \vspace{-0.7em}
 \caption{
   Evolution of the plasmonic response of a sodium nanorod under
   stretching.
   (a) Selected snapshots of the simulated stretching process of the
   nanorod.  Elongation distances $d$ from A to H: 0, 6, 10, 14, 20,
   22, 26, and 28~\AA.
   (b) Longitudinal photoabsorption spectra of the stretched nanorod
   as a function of energy $\hbar \omega$. Solid lines following the
   plasmon modes (see main text for the label definitions) are drawn
   to guide the eye.
   (c) IID maps [see Eq.~\eqref{eq IID}] as a function of energy and
   $z$ coordinate along the main axis of the nanorod.  The maps have
   been smoothed along $z$ to emphasize the gross features of the
   induced density \cite{SupplementaryMaterial}.  Insets show the
   corresponding geometries (right) and spectra (top). Note that in
   panels E, F, and G, the low-energy part of the IID map and spectrum
   have been scaled by the indicated factor.
   \label{fig:spectra}}
\end{figure*}

For our study, we consider a flat-edged sodium nanorod obtained by
extracting 25 atomic layers from the \emph{bcc} lattice along the
$(001)$ direction.  The resulting system, shown in panel A of
Fig.~\ref{fig:spectra}(a), comprises 261 atoms. The lattice constant
was optimized to 4.14~\AA\ by energy minimization of the nanorod (see
below for the numerical details), leading to an approximate diameter
of 13~\AA\ and an aspect ratio of 4. This size is close to the
smallest metallic nanorods produced in recent experiments
\cite{Takahata2014}.  The use of sodium for modeling is not only
computationally convenient because of the single valence electron per
atom needed in its description, it has also been successful in
reproducing many general phenomena of simple and noble metals in
cluster science and plasmonics, both within atomistic
\cite{Zhang2014,Barnett1997,Nakamura1999,Barbry2015,Varas2015} and
continuum approaches
\cite{Zuloaga2009,Ogando2003,Koskinen1995,Marinica2012}.

\nocite{Gao2005,Bernadotte2013,Zuloaga2010,Blochl1994,Ferry2009}

The initial perfect nanorod is adiabatically stretched by pulling
apart the frozen ends in steps of $0.05$~{\AA}, as schematically
indicated in panel A of Fig.~\ref{fig:spectra}(a). After each
elongation step, the atomic positions of the sandwiched central slab
of 51 atoms are allowed to relax.  The geometry relaxation is
performed with density-functional theory
\cite{Hohenberg1964,*Kohn1965} calculations using the \textlcsc{GPAW}
code \cite{Enkovaara2010,*Larsen2009,*Mortensen2005,NoteRelax}.  We
use the Perdew-Burke-Ernzerhof \cite{Perdew1996,*Perdew1997} (PBE)
exchange-correlation (xc) functional. Selected snapshots of the
simulated stretching process are shown in Fig.~\ref{fig:spectra}(a),
labeled from A to H. During the stretching, the break junction
undergoes alternating sequences of elastic deformations and stochastic
rearrangements.  In this process, well-ordered structures can be
formed in the junction, associated to nanowires with stable
\emph{magic radii} \cite{Yanson1999,Ogando2003}, a clear example of
which is observed in panel F. Our simulation is in line with previous
approaches studying the breaking of metallic nanowires
\cite{Nakamura1999,Ogando2003,Jelinek2008,Zugarramurdi2011,Barzilai2013}.

The optical properties of the described stretched nanorod are studied
with time-dependent density-functional theory (TDDFT) \cite{Runge1984}
calculations.  In brief, the time-dependent Kohn-Sham equations are
solved in real time \cite{Yabana1996} with the adiabatic PBE xc
functional as implemented in \textlcsc{GPAW}
\cite{Walter2008,Kuisma2015,NoteTDDFT}.  In this framework, the
plasmons arise from the correlated time evolutions of the Kohn-Sham
electrons in the mean field produced by other electrons.  We consider
only the longitudinal optical response along the main axis ($z$) of
the nanorod. The initial ground-state electronic structure of the
system is perturbed by a weak delta-pulse electric field. The
subsequent computation of the time evolution of the valence electrons
yields the induced charge density $\delta \rho(\V r, t)$, from which
the dynamic polarizability $\alpha(\omega)$ can be derived.  The
response is characterized via the photoabsorption cross section of the
system
\begin{equation}
\sigma_{\text{abs}}(\omega) = \frac{\omega}{c \epsilon_0} \Im[\alpha (\omega)],
\label{eq S}
\end{equation}
where $c$ is the speed of light and $\epsilon_0$ the vacuum
permittivity. The plasmon resonances appearing in the photoabsorption
spectrum can be further analyzed by the imaginary part of the
integrated induced density (IID) defined as
\begin{equation}
\delta \tilde{\rho}(z, \omega)
 = \Im\bigg[\int \delta \rho(\V r, \omega) \, \dee x \dee y \bigg].
\label{eq IID}
\end{equation}
Note that the polarizability can be obtained from the IID via
$\Im[\alpha(\omega)] \propto \int z \, \delta \tilde{\rho}(z, \omega) \, \dee z$.
Thus, the IID serves as a $z$-resolved spectrum summarizing the
spatial distribution of the plasmon modes.

We start our discussion by characterizing the different plasmon modes
found in the calculated spectra shown in Fig.~\ref{fig:spectra}(b) for
the increasing elongation distance $d$.  The breaking of the narrowing
junction after $d=26$~{\AA} divides the spectra into two different
regimes. We first focus on the contact regime ($d \leq 26$~{\AA}),
characterized by the progressive quenching and emergence of several
different plasmon modes.

Initially ($d=0$) the spectrum of the nanorod is governed by a
longitudinal plasmon (LP) mode at 1.5~eV. The dipolar character of
this mode can be recognized in its corresponding IID in panel A of
Fig.~\ref{fig:spectra}(c). Additionally, we observe a faint
higher-order multipolar mode at 2.9~eV (LP2). As the nanorod is
stretched into a connected subnanorod dimer system, the evolved LP
mode is better described as a charge-transfer plasmon (CTP),
consistent with the literature on nanoparticle dimers
\cite{Savage2012,Scholl2013,Barbry2015,Varas2015,Zhang2014,%
Zuloaga2009,Esteban2012,Marinica2012,Liu2013,Song2012,PerezGonzalez2011,%
Duan2012,Song2011,Tan2014,Kulkarni2015,Yan2015}.
The former label, LP, emphasizes the single nanorod character of the
mode. On the other hand, the latter label, CTP, reflects that the mode
can only be supported by the charge flow through the forming
conductive junction between the nanorod ends.  Additionally, the LP2
mode of the single nanorod evolves into a bonding dipolar plasmon
(BDP) in the dimer, both modes having a similar nodal structure in the
IID.  The BDP mode arises from the hybridization
\cite{Nordlander2004,Liu2013} of the dipolar plasmon modes of the two
subnanorods. In contrast to the CTP, the BDP mode exists via
capacitive coupling, also when the junction is removed.  For
increasing $d$, the LP--CTP mode quenches accompanied by a broadening
whereas the initially faint LP2--BDP mode grows, eventually becoming
the dominant resonance in the response.  Both peaks undergo strong
redshifts, consistent with experiments and classical calculations
performed on conductively connected metallic nanoparticles
\cite{Alber2012,Duan2012,Fontana2014,Wen2015}. Thus, the redshifts can
be ascribed to the combined effect of the simultaneous reduction of
the diameter of the junction and the increasing length of the system.

For $d\gtrsim13$~\AA, a second charge-transfer plasmon (CTP2) mode
emerges at $\sim0.3$~eV below the partially overlapping CTP peak. The
CTP2 progressively grows with increasing $d$ and eventually becomes
comparable in strength to the CTP mode.  A careful inspection of the
IID maps in Fig.~\ref{fig:spectra}(c) (panels D--G) reveals subtle
differences between these two charge-transfer modes. While the CTP
mode retains much of its dipolar character with a mostly localized IID
in the far ends of the entire system, the CTP2 has a strong
contribution from both the far and inner ends of the forming
subnanorods.  To further investigate these two modes, we performed
additional calculations with different system sizes and junction
geometries (not shown) suggesting that (i) the CTP2-like charge
oscillation is likely to be dominant for thin junctions and (ii) the
two modes possibly merge in larger systems, in agreement with
classical calculations where a single dominant charge-transfer mode is
observed \cite{Fontana2014,Alber2012}.

Right before breaking, at $d=26$~{\AA}, the stretched nanorod reaches
the interesting limiting case of a monoatomic contact.  In this
configuration, the junction is almost detached from the upper
subnanorod [see panel G in Fig.~\ref{fig:spectra}(a)] and the CTP mode
is strongly suppressed.  In addition to the discussed charge-transfer
plasmon modes, a broad structure of comparable strength is observed at
1.6~eV.  The corresponding IID map (G) in Fig.~\ref{fig:spectra}(c)
reveals the contribution of the junction to the mode. Thus, this
atomic configuration can be viewed as a heterogeneous trimer.

After the breaking of the junction ($d > 26$~\AA), the spectrum
changes abruptly.  We observe that the CTP modes disappear
completely. This is in contrast with the progressive onset of the CTP
modes supported by quantum tunneling when bringing two metallic
nanoparticles into contact \cite{Scholl2013,Savage2012}.  In our
case, the quantum tunneling is rapidly suppressed after the breaking
because of (i) the contraction of the broken junction towards the
subnanorods, thus increasing the vacuum gap, and (ii) the small
available tunneling area in the resulting geometry. Another major
change in the spectrum is the appearance of a second bonding dipolar
mode (BDP2) at 1.3~eV. From the IID map shown in panel H of
Fig.~\ref{fig:spectra}(c) we observe that this mode is associated with
a strong dipolar charge oscillation within the attached remnant
junction and the lower subnanorod, inherited from the weak
pre-breakage feature at 1.6~eV discussed above.

\begin{figure}[h!]
 \centering
 \includegraphics[scale=1]{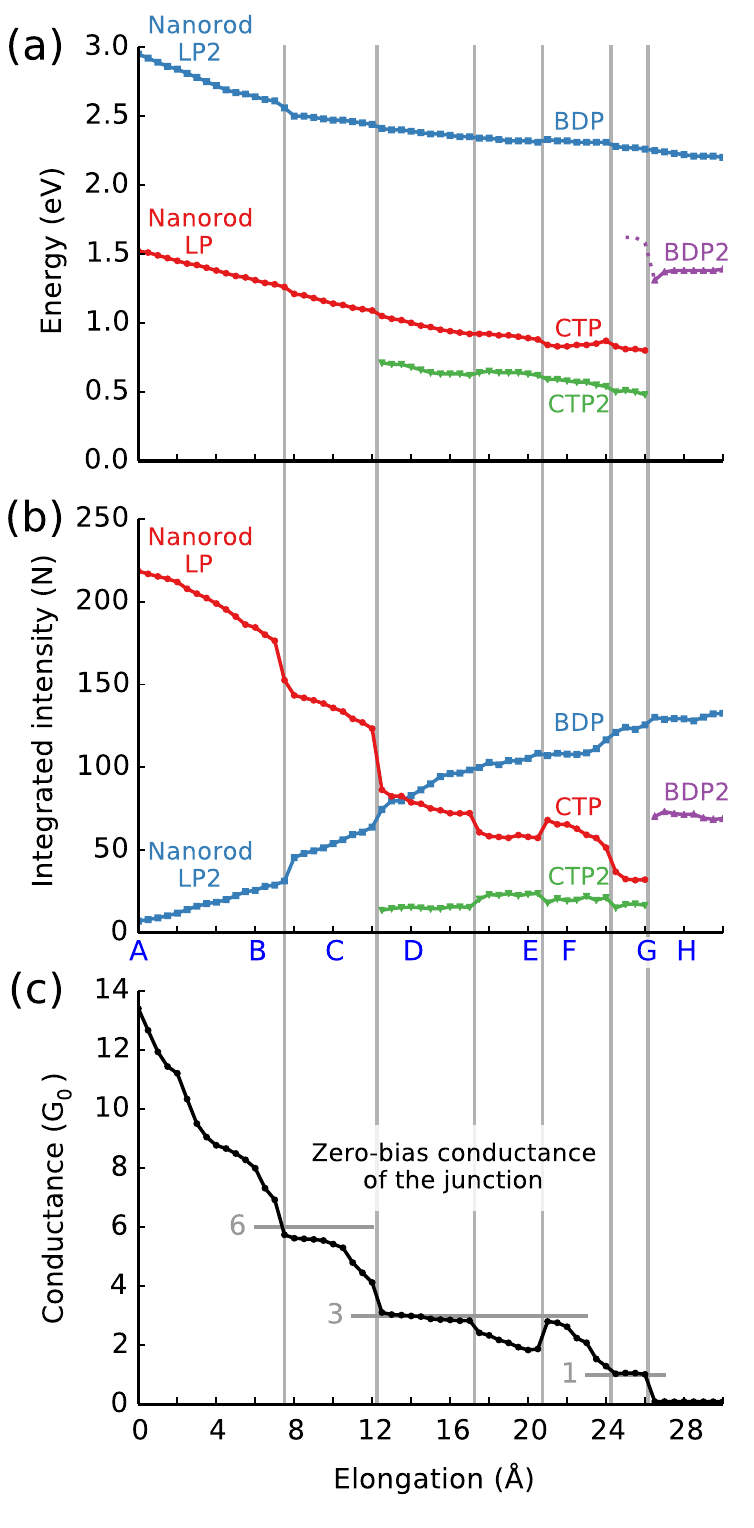}
 \caption{
   Detailed analysis of the plasmon modes during the nanorod
   stretching.
   (a) Peak energy and (b) integrated intensity (the area under the
   peak) of the plasmon modes as a function of the nanorod elongation
   $d$ \cite{NotePeaks}.  The intensities are normalized so that the
   full spectrum integrates to the number of valence electrons
   ($261$).  The line labels correspond to the labeling used in
   Fig.~\ref{fig:spectra}(b) and labels A--H to the geometries shown
   in Fig.~\ref{fig:spectra}(a).  
   (c) Zero-bias conductance of the junction formed during the
   stretching.  Horizontal lines mark values of 1, 3, and 6~$G_0$.
   Vertical lines mark the discontinuities found in the evolution of
   the plasmon modes.}
 \vspace{-0.1em}
 \label{fig:peaks}
\end{figure}

Having characterized the plasmon modes in the system, we proceed to
analyze in detail their evolution. For this aim, we show in
Figs.~\ref{fig:peaks}(a) and \ref{fig:peaks}(b) the extracted energies
and intensities of the plasmon modes as a function of the elongation
\cite{NotePeaks}.  These figures reveal the rich structure in the
evolution of the modes.  The stepped evolution of the different
plasmon modes is particularly remarkable.  This is especially
pronounced for the CTP mode that undergoes strong intensity drops
during stretching at many elongation values. We also observe
discontinuous shifts in the plasmon energies at the same elongations,
most notably at $d=7.5$~\AA\, where the BDP and CTP modes redshift by
0.2~eV and 0.1~eV, respectively.  The discontinuities can be traced
back to the rearrangements in the junction, but such strong intensity
changes can hardly be explained by the geometrical change alone.
Indeed, the discontinuous evolution of the intensity of the CTP mode
is very similar to the quantum features shown by the conductance
curves measured during the breaking of metallic nanowires
\cite{Agrait2003}.

To examine the connection between the observed discontinuities and the
electron transport properties of the junction, we have calculated the
zero-bias conductance of the junction by the non-equilibrium Green's
function technique \cite{Taylor2001} as implemented in \textlcsc{GPAW}
\cite{Chen2012,NoteConductance}. \nocite{Rossi2015Nanoplasmonics} The
conductance results plotted in Fig.~\ref{fig:peaks}(c) show the
well-known features arising in breaking atomic-sized junctions from
the quantization of the conduction channels \cite{Agrait2003}: Upon
stretching, in addition to the gradual evolution of conductance
(reflecting changes in conduction channels' transmission
probabilities), clear jumps of integer multiples of the quantum
conductance unit $G_0=2e^2/h$ take place (due to the closing of
channels).  When these steps are compared with the discontinuities
found in the evolution of the plasmon modes, we observe a nearly
one-to-one match. Thus, we conclude that upon the stretching of the
nanorod, the number of conduction channels supported by the junction
is reduced in a discrete manner \cite{SupplementaryMaterial} after the
atomic rearrangements in the junction; this leads to abrupt changes in
the allowed charge flow between the subnanorods as shown by the
conductance, which is further reflected in the plasmonic response.
This phase-coherent transport effect is particularly pronounced in the
CTP absorption intensity, as expected from the character of the mode.

While the intensity of the CTP mode closely follows the trends shown
by the conductance of the junction, even reflecting the increase of
the conductance at $d\approx 21$~\AA\ (ascribed to enhanced
transmission due to the formation of the well-ordered junction
geometry), the CTP2 mode behaves differently. This is attributed to
the previously described differences in the spatial characteristics of
the two modes: The reduction of the effective cross section of the
junction along with the elongation uncovers the inner surfaces of the
subnanorods necessary to support the CTP2 mode and leads to its
growth.  The BDP mode, also having a strong localization in the inner
ends of the subnanorods, is similarly affected.  Further fine details
in the evolution of the spectra can be attributed to relatively
complex stochastic atomic configurations.

In conclusion, we performed a first-principles study of the plasmonic
response of a stretched metallic nanorod. We found a quantized
evolution of the plasmon modes and established a connection between
the plasmonic discontinuities and the discrete number of conduction
channels supported by the break junction.  Thus, this phenomenon can
be considered the plasmonic counterpart of the quantization of the
conductance in atomic-sized junctions.  The effect is most prominent
in the main charge-transfer plasmon of the considered sodium nanorod.
In noble metal systems the damping due to the d-electrons is weak for
such a low-energy plasmon mode \cite{Sonnichsen2002,Lopez-Lozano2014},
and thus, we expect similar quantized evolution to be observable also
in these materials.  As our findings are based on the formation of a
thin nanocontact, they are most likely to be experimentally
reproducible in a contact breaking situation
\cite{Mennemanteuil2014,Wen2015} of nanoparticle dimers rather than in
a jump to contact one 
\cite{Duan2012,Savage2012,Scholl2013,Barbry2015,Varas2015}.

\begin{acknowledgments}
We thank the Academy of Finland for support through its Centres of
Excellence Programme (2012--2017) under Projects No.\ 251748 and
No.\ 284621. T.\! P.\! R.\ acknowledges financial support from the Vilho,
Yrj\"{o} and Kalle V\"{a}is\"{a}l\"{a} Foundation of the Finnish
Academy of Science and Letters.  We acknowledge computational
resources provided by the Aalto Science-IT project and CSC -- IT
Center for Science Ltd.\ (Espoo, Finland).
\end{acknowledgments}

\emph{Note added.}---A preprint of a related study has recently appeared \cite{Marchesin2015}.

\bibliography{article}

\begin{thebibliography}{65}
\bibitem [{Not({\natexlab{a}})}]{NoteRelax}%
  \BibitemOpen
  \bibinfo {note} {The atom positions were relaxed by a quasi-Newton method.
  Atomic forces were converged to less than $0.01$~eV/\AA. Calculations were
  done within the linear combination of atomic orbitals (LCAO) description
  using the double-$\zeta$ polarized basis set, grid spacing of $0.4$~\AA, and
  vacuum space around the system $\geq\! 6$~\AA.}\BibitemShut {Stop}%
\bibitem [{Not({\natexlab{b}})}]{NoteTDDFT}%
  \BibitemOpen
  \bibinfo {note} {TDDFT calculations were performed on a real-space grid with
  grid spacing of $0.4$~{\AA} and zero boundary conditions. The size of the
  calculation cell was set to $25.6 \times 25.6 \times 96.0$~\AA$^3$ ensuring a
  vacuum space around each geometry $\geq\! 6$~{\AA}. The Hartree potential was
  evaluated in a larger cell of $64 \times 64 \times 128$~\AA$^3$ (vacuum space
  $\geq\! 24$~{\AA}). The propagation time step was $20$~as. A Gaussian folding
  with a full width at half maximum of $0.19$~eV was applied to the
  spectra.}\BibitemShut {Stop}%
\bibitem [{Not({\natexlab{c}})}]{NotePeaks}%
  \BibitemOpen
  \bibinfo {note} {We fitted Voigt profiles to the polarizability spectrum
  $\protect \text{Im}[\alpha (\omega )]$ and extracted the peak parameters of
  the absorption spectrum via Eq.~\protect \textup {\hbox {\mathsurround \z@
  \protect \normalfont (\ignorespaces \ref {eq S}\unskip \@@italiccorr )}}. We
  checked that the obtained trends are robust with respect to the fitting
  procedure.}\BibitemShut {Stop}%
\end{thebibliography}

% Supplemental material
\balancecolsandclearpage


\section*{Supplementary material}

# Supplemental Material for "Quantized evolution of the plasmonic response in a stretched nanorod"

T. P. Rossi,[1,*] A. Zugarramurdi,[1] M. J. Puska,[1] and R. M. Nieminen[1,2]

[1] *COMP Centre of Excellence, Department of Applied Physics, Aalto University School of Science, P.O. Box 11100, FI-00076 Aalto, Finland*

[2] *Dean's Office, Aalto University School of Science, P.O. Box 11000, FI-00076 Aalto, Finland*

[*] E-mail: tuomas.rossi@alumni.aalto.fi

## S1  Smoothing of the integrated induced density maps

For completeness, we show in Fig. S1 the same data as in Fig. 1(c) of the Letter, but without applied smoothing along the $z$ coordinate. The smoothed figure emphasizes the envelope of the induced density that is important for characterizing the plasmon modes. In contrast to the smoothed data, the raw data shows rapid oscillations along the $z$ coordinate. These oscillations can be attributed to size effects and charge screening at the surfaces (Friedel-like oscillations) [1–4].

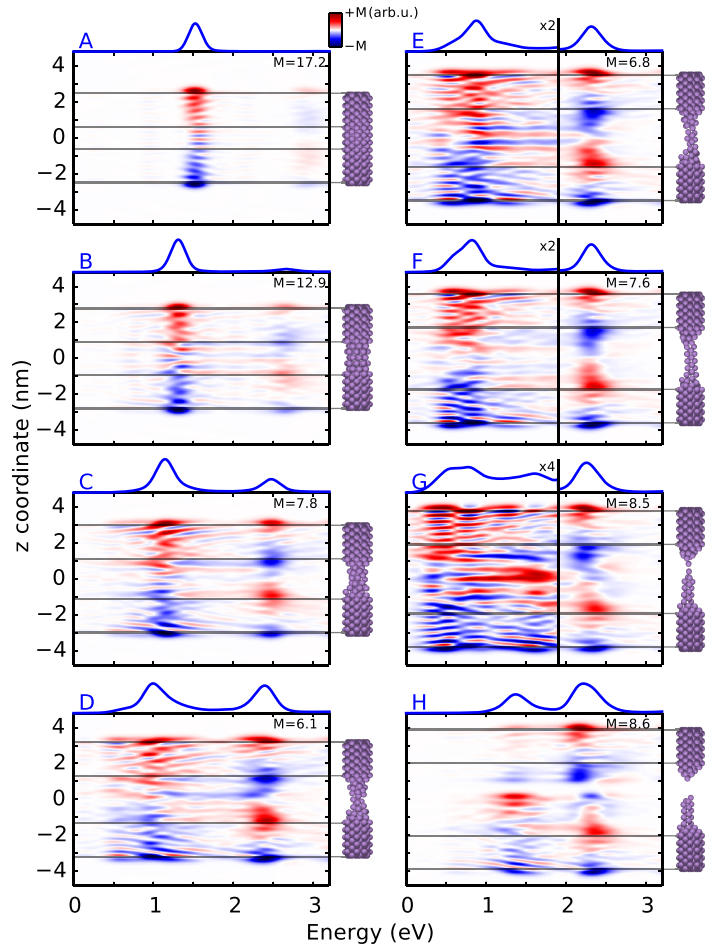

Figure S1: IID maps with no smoothing along the $z$ coordinate.

## S2  Junction electronic structure and transport properties

In this section we provide a more detailed analysis of the junction electronic structure leading to the quantization effects observed in our study. To this end we calculate the junction projected density of states (DOS) and show the quantized electronic structure giving an estimation of the conduction channels contributing to the charge transfer plasmons found in the stretched nanorod.

We define the junction projected DOS (jDOS) as the projection of the density of states onto the junction,

$$g_{\text{junct}}(E) = \sum_i w_i \delta(E - \epsilon_i), \tag{S1}$$

where $w_i$ and $\epsilon_i$ are the weight and eigenenergy, respectively, of the $i$th Kohn-Sham electron state with the wave function $\psi_i(\bm{r})$. The weight $w_i$ is defined as [5]

$$w_i = \int_{z \in [z_0 - \Delta z, z_0 + \Delta z]} |\psi_i(\bm{r})|^2 \mathrm{d}\bm{r} \bigg/ \int |\psi_i(\bm{r})|^2 \mathrm{d}\bm{r}. \tag{S2}$$

The first integral on the RHS of Eq. (S2) is calculated within a slab of width $2\Delta z$ centered at $z_0$, the position of the plane with the smallest junction cross section. Here, $\Delta z$ is chosen as 4 Å and $z_0$ is obtained from the minimum of the $x,y$-integrated ground-state electron density $n^{x,y}(z) = \int \int n(\bm{r}) \mathrm{d}x \mathrm{d}y$, the atomic-scale variations of which have been smoothed along the $z$ direction.

In Fig. S2 we show the calculated jDOS for the geometry F of the stretched nanorod [see Fig. 1(a) in the Letter]. For visualizing the results we have substituted the Dirac delta function $\delta(E - \epsilon_i)$ in Eq. (S1) by a normalized Gaussian function

$$G_\sigma(E - \epsilon_i) = \frac{1}{\sigma\sqrt{2\pi}} e^{-(E-\epsilon_i)^2/2\sigma^2}. \tag{S3}$$

In Fig. S2, the results are shown for smoothing parameter values $\sigma$ of 0.15 eV and 0.03 eV. By considering first the jDOS with $\sigma = 0.15$ eV, we observe two asymmetric peaks with long tails towards higher energies. Their shapes are characteristic for a one-dimensional electron gas (1DEG) with the DOS $g(E) \sim E^{-1/2}$ [7]. This hints that we could model the junction as a idealized cylindrical confining potential of a length $L$ and a diameter $D$, where $L$ is large compared to $D$. The single-electron states in such a potential are characterized by the quantum numbers $(m, n)$ where

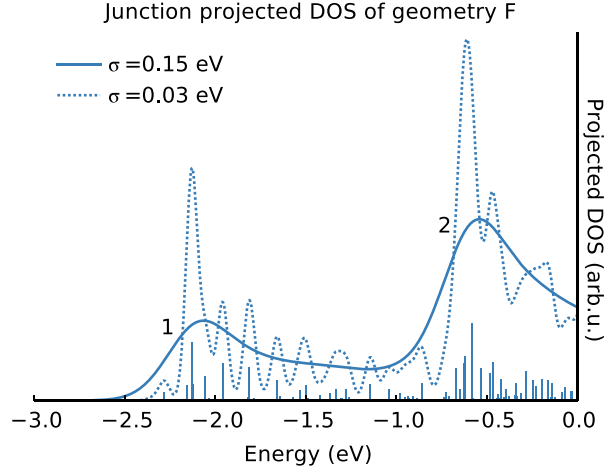

Figure S2: The junction projected density of states of the geometry F (elongation 22 Å). The Fermi level is at 0 eV. Note the characteristic DOS shape of a one-dimensional electron gas. The thin bars indicate the weights $w_i$ of the individual states.

$m = 0, \pm 1, \pm 2, \ldots$ is the angular quantum number and $n = 1, 2, 3, \ldots$ the radial quantum number. Indeed, we can confirm the validity of this idealized picture by visually inspecting the wave functions contributing most to the two labeled peaks in jDOS. In general, the wave functions extend over the whole stretched nanorod, but by focusing at the junction region, we can associate the lower peak (1) to a state corresponding to $(m, n) = (0, 1)$ and the higher peak (2) to two nearly-degenerate states corresponding to $(m, n) = (\pm 1, 1)$. With the increased energy resolution ($\sigma = 0.03$ eV) we can distinguish the discrete states yielding the characteristic DOS shape. By visually inspecting the wave functions of the states at the long tails of the characteristic 1DEG DOS, we can observe that these states correspond to the same $(m, n)$ numbers but have different numbers of nodes along $z$ and/or they are mixed with different states at nanorod ends. Thus, the present metallic junction coupled with the nanorod ends through metal bonding appears to be long enough to support (quantized) one-dimensional electron "bands" over its length. These "bands" serve as conduction channels for the phase-coherent charge transport through the junction. This situation differs from molecular junctions [8, 9] which might have very different electronic structures and coupling mechanisms to the nanoparticles forming the dimer.

In Fig. S3 we show the evolution of the jDOS for the occupied states during the nanorod elongation. We observe clear maxima in the jDOS which are associated to states having stronger density at the junction. These states generally rise in energy as the junction shrinks during the elongation. By visually inspecting wave functions that contribute most to the jDOS peaks, we can associate the states to the following $(m, n)$ states of the cylindrical symmetry:

1) $(m, n) = (0, 1)$ (1 state),
2) $(m, n) = (\pm 1, 1)$ (2 states),
3) $(m, n) = (\pm 2, 1)$ and $(m, n) = (0, 2)$ (close in energy, 3 states in total),
4) $(m, n) = (\pm 3, 1)$ (2 states), and
5) $(m, n) = (\pm 1, 2)$ (2 states).

Note that the states with $m \neq 0$ are not exactly degenerate in the real non-uniform system. The energies of the states closely coincide with those of a uniform cylindrical jellium (Na) nanowire of similar diameter [10]. All in all, this statistical

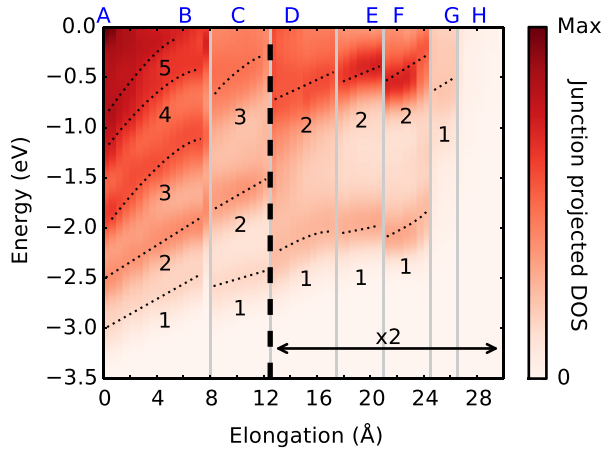

Figure S3: The junction projected density of states as a function of the nanorod elongation [Eqs. (S1)–(S3) with $\sigma = 0.15$ eV]. The Fermi level is at 0 eV. Note that the the DOS has been multiplied by a factor of two for elongations larger than 12 Å. The dotted black lines track the general evolution of the main peaks labeled 1–5. The vertical gray lines and the geometry labels A–H are consistent with Fig. 2 in the Letter.

3

picture shows good agreement with the idealized model presented above in spite of the variations in the exact junction geometry during the stretching.

The jDOS together with the presented model provides direct estimate on the number of open conduction channels supported the junction. By comparing the numbers of "bands" to the calculated conductance [see Fig. 2(c) in the Letter or Fig. S4 below], we observe that the numbers serve as upper values for the conductance in units of $G_0$. In the elongation region denoted by C in Fig. S3, the total number of channels is estimated to be $1 + 2 + 3 = 6$; in the regions D–F $1 + 2 = 3$, and finally, for the single atom contact in the region G, there is one single channel. The atomistic nature and non-uniformity of the junction affects the transmission probability of each conduction channel, and thus the conductance values appear generally below these upper limit estimates.

Next, we compare the transport properties of the junction, determined by the number of conduction channels and their transmission probabilities, to the charge transferred through the junction at the CTP energy. To this end, we have estimated the induced charge at the CTP mode as

$$\delta q(\omega_{\text{CTP}}) = \int_{z \in [-\infty, z_0]} \delta\rho(\bm{r}, \omega_{\text{CTP}}) \, \mathrm{d}\bm{r}, \tag{S4}$$

where $\delta\rho(\bm{r}, \omega_{\text{CTP}})$ is the induced charge density at the CTP energy $\omega_{\text{CTP}}$, and the integration limit $z_0$ is the position of the plane with the smallest junction cross section as discussed above. In other words, the integration is carried over one of the nanorod ends. In Fig. S4 we show the induced charge on top of the calculated zero-bias conductance [see Fig. 2(c) in the Letter]. The agreement between these two quantities is remarkable even though the frequency-independent zero-bias conductance is compared to the light-induced charge. The agreement is clearest after the first discontinuity, after which the thin junction becomes the restrictive factor determining the properties of the CTP mode.

Finally, we briefly note that the induced charge is directly related to the optical linear response of the system discussed in the Letter. The imaginary part of the dynamic polarizability, calculated as the Fourier transform of the time-dependent dipole moment normalized to the perturbing electric field strength, can be estimated as $\text{Im}[\alpha(\omega_{\text{CTP}})] \propto L \, \text{Im}[\delta q(\omega_{\text{CTP}})]$, where $L$ is the effective length of the dipole over which the charge is oscillating at the CTP energy.

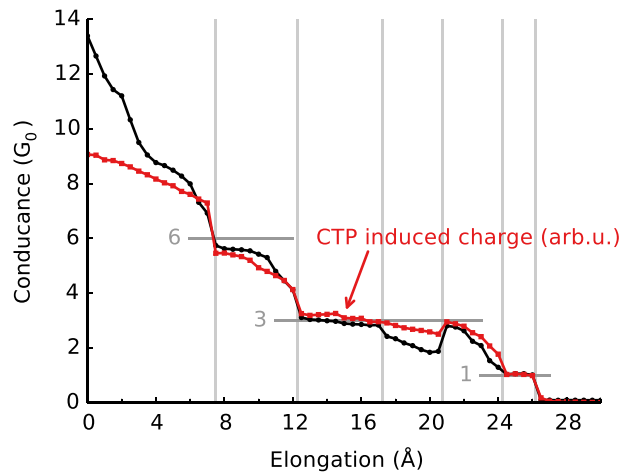

Figure S4: Comparison between the conductance and the induced charge at the CTP mode energy. Black line with circles: zero-bias conductance of the junction during the stretching [from Fig. 2(c) in the Letter]. Red line with squares: the absolute value of the induced charge $|\delta q(\omega_{\text{CTP}})|$ ($\approx \text{Im}[\delta q(\omega_{\text{CTP}})]$) during the stretching. The vertical and horizontal gray lines are consistent with Fig. 2(c) in the Letter.

4



\end{document}